\documentclass[sigconf]{acmart}

\usepackage{booktabs} 
\usepackage{url}

\usepackage{breakurl}

\PassOptionsToPackage{bookmarks=false}{hyperref}
\hypersetup{draft}

\copyrightyear{2018}
\acmYear{2018}
\setcopyright{rightsretained}
\acmConference[MOBILESoft '18]{MOBILESoft '18: 5th IEEE/ACM International Conference on Mobile Software Engineering and Systems }{May 27--28, 2018}{Gothenburg, Sweden}
\acmBooktitle{MOBILESoft '18: MOBILESoft '18: 5th IEEE/ACM International Conference on Mobile Software Engineering and Systems , May 27--28, 2018, Gothenburg, Sweden}\acmDOI{10.1145/3197231.3197234}
\acmISBN{978-1-4503-5712-8/18/05}

\acmDOI{10.1145/3197231.3197234}

\acmPrice{15.00}

\acmSubmissionID{ICSE-CE-MOBILESoft-41}

\begin{document}
\title{MobiCoMonkey - Context Testing of Android Apps}

\author{Amit Seal Ami}
\affiliation{%
  \institution{Institute for Information Technology\\University of Dhaka}
  \city{Dhaka}
}
\email{amit.seal@iit.du.ac.bd}

\author{Md. Mehedi Hasan}
\affiliation{%
  \institution{Institute for Information Technology\\University of Dhaka}
  \city{Dhaka}
}
\email{mehedi.iitdu@gmail.com}

\author{Md. Rayhanur Rahman}
\affiliation{%
  \institution{Institute for Information Technology\\University of Dhaka}
  \streetaddress{Institute for Information Technology\\University of Dhaka}
  \city{Dhaka}
}
\email{rayhan@du.ac.bd}

\author{Kazi Sakib}
\affiliation{%
  \institution{Institute for Information Technology\\University of Dhaka}
  \city{Dhaka}
}
\email{sakib@iit.du.ac.bd}

\begin{abstract}
The functionality of many mobile applications is dependent on various contextual, external factors. Depending on unforeseen scenarios, mobile apps can even malfunction or crash. In this paper, we have introduced  \(MobiCoMonkey\)  - automated tool that allows a developer to test app against custom or auto generated contextual scenarios and help detect possible bugs through the emulator. Moreover, it reports the connection between the bugs and contextual factors so that the bugs can later be reproduced. It utilizes the tools offered by Android SDK and \texttt{logcat} to inject events and capture traces of the app execution.
\end{abstract}

%
%


\keywords{Android, Mobile, Simulation, Contextual Testing, Stress Testing, Emulator, Empirical Software Engineering }

\ccsdesc[300]{Computing methodologies~Simulation tools}
\ccsdesc[300]{Software and its engineering~Integrated and visual development environments}
\ccsdesc[300]{Software and its engineering~Software testing and debugging}
\ccsdesc[300]{Software and its engineering~Empirical software validation}

\maketitle
\section{Introduction}
Performance of Android applications is dependent on various external contextual factors, such as device and OS heterogeneity, network condition, user interaction with external settings and data connection reliability. Due to the differences in contextual scenarios in different geographic regions, apps may not perform well in all conditions. For example, in a network where packet drop is frequent or network disruption is common; an app may lag, show incorrect result or can even crash. This is a severe issue as only 16\% of users return to a bug prone app more than twice \cite{techcrunch-16}. There has been many different testing approaches designed by researchers and industry organizations to find bugs and test apps, such as \cite{Moran2016}, \cite{GuiRippingAmalFitano}, \cite{Amalfitano2011},  \cite{Moran:2017:OBR:3104086.3104133}, \cite{Baek2016AutomatedCriteria}, \cite{Ravindranath2014}, \cite{Ye2013}, \cite{Sasnauskas2014},  \cite{google-monkey}, \cite{xamarin-url},  \cite{10.5120/ijca2017915210} and \cite{Liang2014}. However, as found by Poshyvanyk et al. \cite{Poshyvanyk2017} the approaches designed by researchers and industry organizations alike are seldom used for various reasons; such as lack of exposure, applicability under certain conditions as  well as incompatibility with existing Standard Development Kit and other third party tools. Another problem of these approaches is lack of reproducibility of event sequences. 

To aid the testing process specially considering the above problems and the contextual factors of mobile applications, we present a lightweight, non obstructive tool called \(MobiCoMonkey\) (\textbf{Mobi}le \textbf{Co}ntextual \textbf{Monkey})\cite{lordamit-monkey} through this work. It allows developers and testers to test an app in different contextual scenarios. Provided a \textit{config} file, it allows user to test an app while executing a set of contextual events either automatically generated or predefined by user. It extracts necessary information from the android app installer (\textit
{apk}) and produces an insightful report that allows one to trace an error or warning back to the relevant contextual conditions. It is capable of working without modifying standard Android SDK and any other third party tools. 

\section{Background Study and Related Works}
As shown by Muccini et al.  \cite{Muccini:2012:STM:2663608.2663615}, mobile apps are different from traditional software and require specialized, new testing techniques. As a result, many different works were done including augmenting existing approaches and introducing novel approaches of mobile app testing.  Several state of the art works focus on providing contextual testing as a cloud service \cite{Liang2014}, testing as a service \cite{google-firebase}, generating different input for Android apps ~\cite{Machiry2013Dynodroid:Apps}, recording input events to later replay the same events to reproduce bugs \cite{reran6606553}, generating automatic GUI testing~\cite{Amalfitano2011}, on device bug reporting for android applications~\cite{Moran:2017:OBR:3104086.3104133}, generating contextual events for stress testing Android Apps ~\cite{10.5120/ijca2017915210}, intent fuzzing ~\cite{Sasnauskas2014} and, automatic discovering, reporting and reproducing mobile app crashes through static and dynamic analysis \cite{Moran2016}. However, as surveyed in  \cite{howdodeveloperstestandroidapps} by Vasquez et al., most of these approaches are not used in industry for various reasons; such as lack of applicability, heavy overhead for integrating in existing approach, and  incompatibility with existing tools. Further, the author made several recommendations for future testing approaches such as automatic test cases that evolve with app, has low overhead to integrate with existing practices, and better trace-ability between test cases and features. 
\(MobiCoMonkey\) focuses on being a low overhead testing tool that utilizes non modified Android SDK and interacts with the app in a non obstructive manner. Therefore, it is possible to utilize it with other existing approaches. Further, it offers trace-ability between the injected contextual events and the resulting logs generated within the activities of the app. This allows the developer to zero-in on the possibly bug prone features. 
\section{The \(MobiCoMonkey\) Tool}
\begin{table}
\caption{Event Types and Possible Values of MobiCoMonkey based on Android Guidelines~\cite{google-settings-global}}
\label{table-contextual_events} %
\begin{tabular}{p{0.8cm}p{1.2cm}p{1cm}p{1cm}p{1.3cm}p{1.2cm}}
    \toprule
GSM & Network & \multicolumn{2}{c}{Network} & Key & User \\
Pro-file & Delay &\multicolumn{2}{c}{Status} & Events & Rotation \\
    \midrule
0  & GSM  & GSM  & HSDPA & Alphabets & Portrait \\

1  & EDGE  & HSCSD  & LTE  & Numerals & landscape \\

2  & UMTS  & GPRS  & EVDO  & Symbols & Reverse Portrait \\

3  & None  & UMTS  & FULL  & Back space & Reverse Landscape \\

4  &  & EDGE  &  & Delete & \\
\bottomrule
\end{tabular}
\end{table}

\(MobiCoMonkey\) consists of several independent, configurable components. An overview of the components are discussed in subsection \ref{subsection0}. Any developer can download it from \texttt{GitHub} \cite{lordamit-monkey} and use \textit{python3.6}, \texttt{JavaFX} and \texttt{pip} to utilize its various components. Underneath, it utilizes various high level and low level elements of \textit{Android SDK}, such as \texttt{adb}, \texttt{emulator}, \texttt{shell}, \texttt{telnet} and \texttt{aapt}. MobiCoMonkey is capable of injecting several low level contextual factors, which are: GSM Profile, Network Delay and Network Status. Additionally, it is capable of changing Rotation View, inserting Key Press events, and toggling Airplane Mode. The categories and values of currently supported events are provided in Table \ref{table-contextual_events}. 
\begin{figure}
\centerline{\includegraphics[width=8cm]{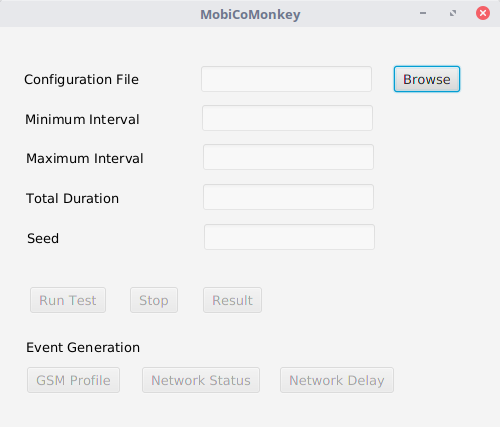}}
\caption{MobiCoMonkey GUI}
\label{figure:gui}
\end{figure}
\begin{figure}
\centerline{\includegraphics[width=8cm]{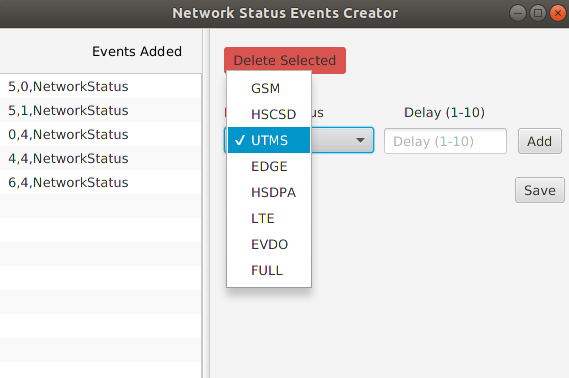}}
\caption{Network Status Events Creator}
\label{figure:generator}
\end{figure}
It should be clarified that contextual network condition variation is only possible in emulated devices. Enabling support for such in actual android devices require high overhead modification of the Android OS and rooting. Therefore, \(MobiCoMonkey\) supports only android emulators as of now. 

\subsection{Components}
\label{subsection0}
The internal components of \(MobiCoMonkey\) are created using Python 3.6. These components are accessed by the \(MobiCoMonkey\) GUI to perform various types of actions. Due to various user configurable settings, \(MobiCoMonkey\)'s nature is quite flexible. The internal components are discussed as follows.

\begin{figure*}[h]
\centerline{\includegraphics[width=12cm]{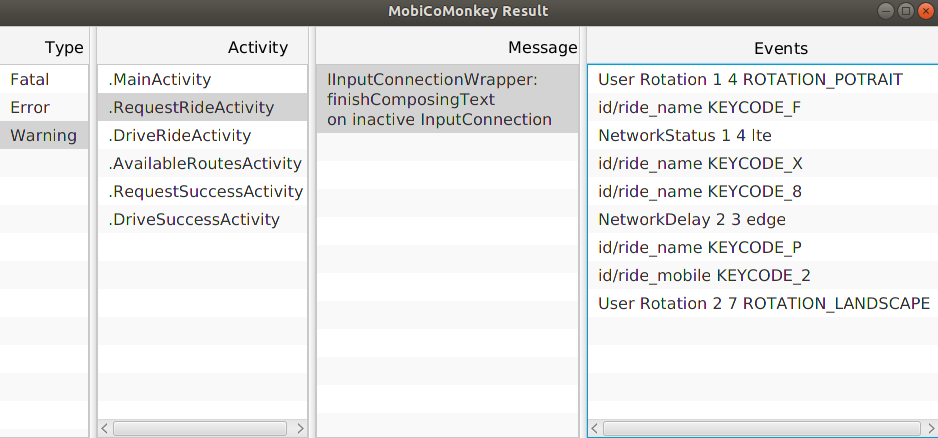}}
\caption{MobiCoMonkey Result Viewer}
\label{figure:result}
\end{figure*}


\subsubsection{App Manager:}
The internal App Manager takes care of installing and uninstalling the app as required. It also extracts permissions requested by the app to determine which \(MobiCoMonkey\) supported contextual factors will be applicable for contextual testing. Additionally, it extracts UI components of an activity by utilizing \texttt{uiautomator dump} when requested. To get the full list of User Interface (UI) elements, it scrolls from top to bottom of the screen until no new UI elements are found in the same activity. As a result, it becomes possible to create mapping of UI elements to each activity regardless of the screen size and view mode.

\subsubsection{Contextual Scenario Generator:}
The contextual scenario generator is utilized when user defined scenarios are not provided for testing. Based on permissions, it generates random contextual scenarios based on a user provided seed value. Accordingly, the same scenarios can always be generated by utilizing the seed value. While generating, it considers minimum interval and maximum interval between contextual factor changes based on user configuration, and the total duration of each scenario. For example, if the minimum and maximum intervals are configured to be 5 and 12, two Network Status related events it may generate are:
\begin{verbatim}
NetworkStatus 0 5 hscsd
NetworkStatus 1 10 lte
\end{verbatim}
Each line represents one contextual event. The first value describes the type of contextual event, the second value is the index, the third value is the interval after which the next Network Status event will be applied and the fourth value is the current contextual event to be applied. The combined duration of all contextual events for a particular type matches the total duration of scenario. All the other contextual scenarios are produced similarly. 
\subsubsection{Executor:}
This can utilize various approaches for contextual testing. For example, it can run in guided approach, where it will run only selected activity screens of an app. Otherwise, it will iterate through all the activities of the provided app. Before executing any contextual events from the scenario, it starts \texttt{logcat} service in the Emulator with a clean state. Next, for each activity it runs the contextual scenario events. Consequently, each executed event are saved in a log file for later analysis. To illustrate, a sample log can be: 
\begin{verbatim}
03-19 00:36:35 NetworkStatus 0 8 lte
03-19 00:36:43 NetworkStatus 1 5 gsm
03-19 00:37:05 UserRotation 0 8 ROTATION_REVERSE_POTRAIT
\end{verbatim}
The generated log file follows a structure similar to logs generated by logcat\cite{google-logcat}. The executor can also additionally run an optional text input field testing thread. Utilizing a top-down approach for each activity, it inputs random text characters in text input fields. Additionally, it stores progress of text fields covered while doing so. Therefore, even if the view is suddenly rotated, it can continue inserting text input sequentially. The executor furthermore watches for fatal status in case of app crash. In such case, it stops execution and saves available logs. 

\subsubsection{Log Analyzer:}
It combines logs generated by the logcat emulator and \(MobiCoMonkey\) executor. The logs are sorted based on temporal ascending order. From the logcat generated log, it extracts the logs of the app at \texttt{Warning}, \texttt{Error} and \texttt{Fatal} level. Generally, these logs are produced by either the system and/or the developer as part of debugging. \texttt{Warning} is defined as a non-obstructive malfunction which takes place when something does not execute properly. Next, \texttt{Failure} or \texttt{Error} is found when an internal activity crashes, while the app resumes execution. For example, \texttt{android.view.WindowLeaked} is found when a dialog is dismissed improperly. Lastly, \texttt{Fatal} is found when the app crashes fatally and forfeits execution. 

\subsubsection{GUI Components}
For ease of use, several Graphical User Interface components are prepared as part of MobiCoMonkey using \texttt{JavaFX}. These components are namely MobiCoMonkey GUI, Contextual Scenario Creator,  and Result Viewer as shown in Figure \ref{figure:gui}, \ref{figure:generator} and \ref{figure:result} respectively. 

Activities related to utilizing \(MobiCoMonkey\) can be categorized to three segments, which are Setup \& Configuration, Execution, and Result Analysis. These are described in sub sections \ref{subsection1}, \ref{subsection2}, and \ref{subsection3} respectively.

\subsection{Setup \& Configuration }
\label{subsection1}
To start, a developer has to download and follow instructions from the \texttt{GitHub} \cite{lordamit-monkey} repository to initially set configuration values of \(MobiCoMonkey\) in a \textit{config} file. It is assumed that the user already has Android SDK, Android Virtual Device and Android Emulator installed in system. The configuration file is utilized by the \(MobiCoMonkey\) GUI after user browses and selects the address of the \textit{config} file. Next, the user can utilize the \textit{Event Generator} Window to create a custom scenario of Contextual Events. For each entry of contextual events, the user has to insert the type of event, the nature of the event and the duration of the event. For example, if the user wants to add a Network Status of GSM stage, he simply has to select it from the drop-down menu and input the delay before the next event. While saving the file, the generator will convert it into a Comma Separated Value(CSV) format that can be easily read by the \(MobiCoMonkey\) \textit{Executor} module. The user can also simply ignore the Event Generator module and click on `Run Test`. 
In such case, the executor module will generate contextual scenarios based on a user defined seed value automatically and will inject those for a prefixed amount of time. 

\subsection{Execution}
\label{subsection2}
After clicking on \textit{Run Test} at first, the app manager module will check for existing installation of the app installer file (\textit{apk}) after initiating the Emulator and will install a fresh version if necessary.  Second, the app manager will extract several meta-data related to the app, such as requested permissions and name of activities included in app. Based on extracted permissions, the executor will filter the events in the provided scenario. For example, an app that does not request permission to access the Internet or any data network might not be affected by a sudden contextual change in Network Status or Network Speed. Third, based on the extracted names of activities, the executor will extract further information related to User Interface, such as available text input fields. As a result, it is possible to traverse through activities in a top-down approach, while inserting textual elements in each input fields. 

During execution, it creates detailed log similar to  \textit{logcat}\cite{google-logcat} \textit{detail format}. This log contains information related to android activities, UI components, and injected events. Additionally, it also utilizes the Android internal \textit{logcat} to monitor the errors, warnings and fatal errors. All these information are stored continuously to prevent loss of data. Furthermore, in case of a fatal error it immediately stops execution and stops collecting unnecessary data. 

\subsection{Result Viewing}
\label{subsection3}
Using the \textit{MobiCoMonkey Result Viewer} as shown in Figure \ref{figure:result}, the information are categorized to \texttt{Warning}, \texttt{Error}, and \texttt{Fatal} levels. 
It then allows further filtering to android activities screens so that user is able to view details only from the necessary activities. By navigating to the appropriate activity, the user is able to view the events that occurred during testing. With it, the time-based adjacent, injected contextual events are provided so that user may deduce whether the derived execution event is influenced by the injected contextual events. 
\section{Conclusion}
The strength of \(MobiCoMonkey\) lies in its simple, modular approach for handling different tasks of contextual mobile testing. An inexperienced user without any idea about contextual testing can use it similar to \textit{Monkey}, where it injects random contextual events to an app. On the other hand, an experienced user can design custom contextual scenarios while utilizing any other kind of testing, such as UI testing or Unit testing.  Furthermore, it allows the user to utilize it without modifying any existing tools in use or modifying the applications. As a result, it can be integrated as part of agile software methodology with very little overhead. In future, \(MobiCoMonkey\) can be extended for further analysis of app behavior so that causation and correlation can be derived from injected contextual events and observed behavior of the app. Additional contextual factors, such as GPS, Accelerometer and Bluetooth can be considered for future scope extension.
\bibliographystyle{ACM-Reference-Format}

\bibliography{mobisoft}

\end{document}